\documentclass[letter]{aa}  
\usepackage{natbib,hyperref}
\usepackage{CJK}
\usepackage{graphicx}
\usepackage{txfonts}

{\newif\ifnotend
\notendtrue
\def\veclist{ABCDEFGHIJKLMNOPQRSTUVWXYZabcdefghijklmnopqrstuvwxyz.}
\def\top#1#2.{#1}
\def\tail#1#2.{#2.}
\loop\expandafter\xdef\csname v\expandafter\top\veclist\endcsname%
{{\noexpand\bf\expandafter\top\veclist}}
\edef\veclist{\expandafter\tail\veclist}
\if\veclist.\notendfalse\fi\ifnotend\repeat}
{\newif\ifnotend
\notendtrue
\def\callist{ABCDEFGHIJKLMNOPQRSTUVWXYZ.}
\def\top#1#2.{#1}
\def\tail#1#2.{#2.}
\loop\expandafter\xdef\csname c\expandafter\top\callist\endcsname%
{{\noexpand\cal\expandafter\top\callist}}
\edef\callist{\expandafter\tail\callist}
\if\callist.\notendfalse\fi\ifnotend\repeat}

\def\Vc{v_{\rm c}}

\def\apjs{ApJS}
\def\apjl{ApJL}

\def\aap{A\&A}

\def\Gyr{\,\mathrm{Gyr}}

\def\kpc{\,\mathrm{kpc}}

\def\kms{\,\mathrm{km\,s}^{-1}}

\def\msun{\,{\rm M}_\odot}

\def\eg{{ e.g.,\ }}
\def\sun{\odot}

\newcommand{\Gaia}{{\emph{Gaia}}}

\newcommand*\samethanks[1][\value{footnote}]{\footnotemark[#1]}

\newcommand{\gkai}[1]{\begin{CJK*}{UTF8}{gkai}\raisebox{.1em}{(}#1\raisebox{.1em}{)}\end{CJK*}}

\renewcommand{\[}{\begin{equation}}
\renewcommand{\]}{\end{equation}}
\def\i{{\rm i}}

\def\np{{ 2290\ }}
\def\nd{{ 284\ }}
\makeatletter

\begin{document}

   \title{Could very low-metallicity stars with rotation-dominated orbits have been shepherded by the bar?}

   % \subtitle{}

   \author{Zhen Yuan \gkai{袁珍}\inst{1}\fnmsep\thanks{\email{zhen.yuan@astro.unistra.fr;\\chengdong.li@astro.unistra.fr}}
	\and Chengdong Li \gkai{李承东}\inst{1}\fnmsep\samethanks[1]
    \and Nicolas F. Martin \inst{1,2}
    \and Giacomo Monari\inst{1}  
    \and Benoit Famaey \inst{1} 
    \and Arnaud Siebert \inst{1}
    \and Anke Ardern-Arentsen \inst{3}
    \and Federico Sestito\inst{4}
    \and Guillaume F. Thomas\inst{5,6}    
    \and Vanessa Hill\inst{7}   
    \and Rodrigo A. Ibata\inst{1}  
    \and Georges Kordopatis\inst{7}  
    \and Else Starkenburg \inst{8}  
    \and Akshara Viswanathan\inst{8}        
 }

    \institute{Universit\'e de Strasbourg, CNRS, Observatoire astronomique de Strasbourg, UMR 7550, F-67000 Strasbourg, France
         \and
         Max-Planck-Institut f\"{u}r Astronomie, K\"{o}nigstuhl 17, D-69117 Heidelberg, Germany
         \and
         Institute of Astronomy, University of Cambridge, Madingley Road, Cambridge CB3 0HA, UK
         \and
	     Dept. of Physics and Astronomy, University of Victoria, P.O. Box 3055, STN CSC, Victoria BC V8W 3P6, Canada 
      	 \and
	    Instituto de Astrof{\'\i}sica de Canarias, E-38205 La Laguna, Tenerife, Spain
	    \and
	    Universidad de La Laguna, Dept. Astrof{\'\i}sica, E-38206 La Laguna, Tenerife, Spain
         \and
         Universit\'e C\^ote d'Azur, Observatoire de la C\^ote d'Azur, CNRS, Laboratoire Lagrange, Nice, France
        \and
         Kapteyn Astronomical Institute, University of Groningen, Landleven 12, 9747 AD Groningen, The Netherlands
 }

% Don't change these lines
\date{Received XXX; accepted XXX}

    \abstract{
    The most metal-poor stars (\eg [Fe/H] $\leq-2.5$) are the ancient fossils from the early assembly epoch of our Galaxy, very likely before the formation of the thick disc. Recent studies have shown that a non-negligible fraction of them have prograde planar orbits, which makes their origin a puzzle. It has been suggested that a later-formed rotating bar could have driven these old stars from the inner Galaxy outward, and transformed their orbits to be more rotation-dominated. However, it is not clear if this mechanism can explain these stars as observed in the solar neighborhood. In this paper, we explore the possibility of this scenario by tracing these stars backwards in an axisymmetric Milky Way potential with a bar perturber. We integrate their orbits backward for 6 Gyr under two bar models: one with a constant pattern speed and another one with a decelerating speed. Our experiments show that, under the constantly-rotating bar model, the stars of interest are little affected by the bar and cannot have been shepherded from a spheroidal inner Milky Way to their current orbits. In the extreme case of a rapidly decelerating bar, some of the very metal-poor stars on planar and prograde orbits can be brought from the inner Milky Way, but $\sim90\%$ of them were nevertheless already rotation-dominated ($J_{\phi}$ $\geq$ 1000 km s$^{-1}$ kpc) 6 Gyr ago. The chance of these stars having started with spheroid-like orbits with small rotation ($J_{\phi}$ $\lesssim$ 600 km s$^{-1}$ kpc) is very low ($<$ 3$\%$). We therefore conclude that, within the solar neighborhood, the bar is unlikely to have shepherded a significant fraction of inner Galaxy spheroid stars to produce the overdensity of stars on prograde, planar orbits that is observed today.
    }


  \maketitle

%%%%%%%%%%%%%%%%% BODY OF PAPER %%%%%%%%%%%%%%%%%%

\section{Introduction}
\label{sec:intro}

Stars with [Fe/H] $\leq -2.5$ were born in the ancient Universe when baryons started to assemble into stars and galaxies \citep[see\eg][]{beers05,frebel10}. These objects are extremely rare because the Galaxy gets enriched above [Fe/H]$ = -2.5$ quickly after the Big Bang, typically within 1 Gyr (equivalent to $z$ $\sim$ 5). The gas from which they form received metals from a handful of earlier supernova explosions at the epoch when the interstellar medium is not well mixed \citep[see \eg][]{argast00} and probably even before the thick disc of our Galaxy gets built up \citep[see \eg][]{gallart19,xiang22}.

Thanks to a variety of spectroscopic and photometric surveys and their follow-up studies that are dedicated to searches for low-metallicity stars \citep[see \eg][]{beers87, christlieb08, strakenburg17, wolf18, lihn18, aguado19}, there are more than two thousand stars with spectroscopic metallicities below $-2.5$. This number keeps increasing with on-going and upcoming spectroscopic surveys, such as the Milky Way Survey from the Dark Energy Spectroscopic Instrument \citep[DESI][]{cooper23} and the WEAVE survey \citep{jin2023}. Most of these stars are nearby and bright and have accurate parallax measurements from \Gaia\ \citep{edr3astro}, which means that their full 6D kinematic information is available. We are therefore able to study their orbital properties that record the dynamical memories of their origins. 

Old and very low-metallicity stars are mostly expected to be the debris from ancient accretion events, which we could expect to naturally produce an isotropic halo distribution in angular momentum if they were accreted on random orbits, have similar masses, and none of them are predominant. However, in the very low-metallicity sample from the LAMOST and Pristine surveys \citep{lihn18, aguado19}, there is a significant asymmetry between the retrograde and prograde planar stars \citep{sestito20}, which is also seen in the ESO "First Stars" program results \citep{dimatteo20} and in the Hamburg/ESO Survey \citep{carollo23}. A population of several hundred stars are rotation-dominated and prograde, including a few ultra metal-poor stars (UMP; [Fe/H] $\leq$ $-$4) that \citet{sestito19} showed have orbits close to the solar orbit. Given their low-metallicity nature, these stars are expected to have been formed much earlier than the disc. Where these very low-metallicity prograde stars come from remains a puzzle. 

\citet{sestito19, sestito20} discussed three possible origin scenarios for these stars: they were (i) accreted from small satellites with specific orbits through minor mergers during the life of the Milky Way; (ii) brought in during the early assembly of the proto-Milky Way disc; (iii) formed in-situ from pockets of pristine gas at early times pushed into the solar neighbourhood, probably through interactions with the Milky Way (MW) bar and its spiral arms \citep{minchev10}. Similar to the migration mechanism discussed in (iii), \citet{dillamore23} proposed a fourth scenario: (iv) halo (pressure-supported) stars, originally in the inner Galaxy, that gained rotation and moved outwards due to the bar resonances.

Among these four possibilities, the exploration of high-resolution cosmological simulations, such as NIHAO-UHD and FIRE, suggest that the in-situ formation from pockets of pristine gas in a thin disk is ruled out \citep{sestito21, santistevan21}. Of the remaining three, probably only the last scenario is straightforward and can be tested by simple experiments. The inner Galaxy is the reservoir of very old and very metal-poor (VMP) stars that is predicted from hydro-dynamical simulations \citep[see \eg][]{strakenburg17,el-badry18} and is seen by the Extremely Metal-poor BuLge stars with AAOmega survey \citep[EMBLA,][]{howes16} and the Pristine Inner Galaxy Survey \citep[PIGS,][]{arentsen20a, arentsen20b} as well as by \Gaia\ \citep{rix22, yao23, martin23}. The very low-metallicity stars ([Fe/H] $\leq -2.5$) of the inner Galaxy are even more metal-poor than $Aurora$ ([Fe/H] $\gtrsim$ $-$2), considered as an in-situ component according to its dynamical properties and chemical features (Al, N) from \citet{belokurov22, belokurov23}. These stars probably belong to the proto-Milky Way comprised of either one (\eg $Aurora$) or, as suggested from zoom-in MW-like simulations, a handful of early accreted systems \citep{horta23}, or many low-mass, now-merged satellites \citep{el-badry18}.

Nevertheless, we could speculate that the oldest stellar populations in the inner Galaxy can be a significant source of the low-metallicity planar stars observed at the present day in the solar neighborhood, as suggested by \citet{dillamore23}. In this letter, we explore this scenario by tracing backwards the current sample of observed very low-metallicity prograde planar stars under the perturbation of a rotating bar. In our experiment, the bar is designed to have either a constant or a decreasing pattern speed and we examine the possibility that the observed stars moved from the inner Galaxies under the influence of these two bar models. We describe the very low-metallicity spectroscopic sample we rely on in Sec.~\ref{sec:data}. The setup of the MW model with the two distinct bar models with different pattern speeds is explained in Sec.\ref{sec:mod}. Finally, the results from the (backwards) orbital integration of the sample are shown in Sec.\ref{sec:res} and these results are discussed in Sec.\ref{sec:dis}.

% In this work, we explore if the bar can empower these stars with strong rotation and drive them outwards.  
\section{DATA}{\label{sec:data}}

In this work, the very low-metallicity sample is selected in a similar way to that of \citet{sestito20}. We combine the LAMOST DR3 VMP catalog \citep{lihn18} with the modified metallicity from \citet{yuan20}, the Pristine sample from \citet{aguado19, sestito20} and the UMP sample from \citet{sestito19}. A simple metallicity cut, [Fe/H]$\leq-$2.5, yields a parent sample of \np stars to start with.

We follow a Bayesian approach to derive distances by combining the photometric and astrometric information \citep{sestito19}. For the prior, we assume that the distribution of the low-metallicity stars follows a halo profile, specifically, the RR Lyrae density profile, $\rho(r)\propto r^{-3.4}$ from \citet{hernitschek18}. We simplify the method by computing the probability distribution function (PDF) as a function of the distance with logarithmic bin size. The majority of stars in our sample are within $5\kpc$ of the Sun.

Since all the stars in our parent sample have spectra taken by telescopes in the Northern hemisphere, we are able to use the spectroscopic stellar parameters and correct their extinction values using the 3D dust map from \citet{green19}. This approach requires distance estimates in the first place. Therefore, we first adopt the 2D dust map from \citet{schlegel98} to provide an estimated distance as the input to the 3D dust map \citep{green19}, which gives a new distance estimate iteratively. In the application of the extinction correction, we use the coefficients derived by \citet{martin23} that depend on the stellar parameters ($T_{\rm{eff}}$, log $g$, [Fe/H]). We feed these parameters, as listed in the survey catalogues and obtained from the spectra, to the formula \citep[equation~2 of][]{martin23}. The radial velocities have several sources from specific spectroscopic analyses \citep{sestito19,sestito20,lihn18} as well as from the $Gaia$ Radial Velocity Sample RVS \citep{dr3rvs}. In cases with multiple radial velocity measurements, we keep the measurement with the smallest uncertainties.

With the 6D kinematic measurements in hand, we are able to compute the orbital parameters for the low-metallicity parent sample using AGAMA \citep{agama}, with the MW potential from \citet{mcmillan17}. In particular, we calculate the actions of the stars ($J_{\phi}$, $J_{\rm r}$, $J_{\rm z}$). There are \nd prograde planar stars in the selection box used by \citet{sestito20}: 0.5 $\leq$ $J_{\phi}/J_{\phi,\sun}$ $\leq$ 1.0, 0.5 $\leq$ $J_{z}/J_{z,\sun}$ $\leq$ 1250, with $J_{\phi,\sun}$ = 2009.92 $\kms\kpc$ and $J_{z,\sun}$ = 0.35 $\kms\kpc$. For each star in the very low-metallicity prograde planar sample, we draw a sample of 500 realizations based on the uncertainties of their 6D kinematic information. Specifically, the distance is sampled from the posterior probability distribution, and the radial velocities are sampled from a Gaussian distribution according to their measured uncertainties. Then stars are sampled in the ($\alpha$, $\delta$, $\mu_{\alpha}$, $\mu_{\delta}$) space after taking into account the covariance matrix \citep{edr3astro}. We now have the final sample of \nd $\times$ 500 particles as the input sample for the backwards orbit integration procedure.

\section{Models}{\label{sec:mod}}

The potential used in this work is made of two components: the axisymmetric background and the non-axisymmetric perturbation. The axisymmetric background potential is constructed through a series of distribution-function-based models containing a dark halo, a stellar halo, a bulge, and stellar discs that are self-consistent with \texttt{AGAMA} \citep{agama}. The distribution function (DF) of each component is a specified function $f(\vJ)$ of the action integrals. In addition, a gas disc which is not included in the DF model is added to derive the total potential of the Galaxy. The detailed setup of this MW self-consistent model and its various predictions can be found in \citet[]{Binney2023d,Binney2023c}. It should be noted that, although the DF-based modelling method is similar, there are some differences between this work and \citet[]{Binney2023d,Binney2023c}. Here, we directly use the self-consistent model implemented in \texttt{AGAMA} with a spheroidal bulge and two quasi-isothermal discs. The example of this code can be seen on-line\footnote{https://github.com/GalacticDynamics-Oxford/Agama/blob/master/py/example$\_$self$\_$consistent$\_$model.py} with the initial parameters for the model\footnote{https://github.com/GalacticDynamics-Oxford/Agama/blob/master/data/SCM.ini}.

The non-axisymmetric perturbations for the Galaxy include two parts : a central bar and spiral arms. We choose two kinds of bar models in this work, a steadily rotating bar and a decelerating one. Both models are discribed in detail in \citet{Li2023a}. We put the bar in the simulation at $t= -6 \Gyr$, let it evolve to the present time at $t= 0 \Gyr$, and integrate orbits backward for each particle from the input sample with the trajectories stored every 0.02 $\Gyr$. The timescale of the simulation (6 Gyr) is chosen to coincide with the estimated formation epoch of the bar \citep[6--8\,$\Gyr$;][]{Wylie2022,Sanders2023}.

The steadily rotating bar, with pattern speed $\Omega_{\rm{b}}$, is modelled following \citet{Chiba2022} as
\begin{equation}
    \Phi_{\rm{b}}(r,\theta,\phi,t)\,=\,\Phi_{\rm{br}}(r)\sin^2{\theta}\cos{m(\phi-\Omega_{\rm{b}} t-\phi_{\rm{b}})},
    \label{eq:phib}
\end{equation}
where $(r,\theta,\phi)$ are the spherical coordinates. We only consider the $m=2$ quadrupole term in this work. The radial dependence of the bar potential, $\Phi_{\rm{br}}(r)$, is
\begin{equation}
    \Phi_{\rm{br}}(r)\,=\,-\frac{A\,\Vc^{2}}{2}\,\bigg(\frac{r}{r_{\rm{CR}}}\bigg)^{2}\,\bigg(\frac{b+1}{b+r/r_{\rm{CR}}}\bigg)^{5},
    \label{eq:phibr}
\end{equation}
where $A$ measures the strength of the bar, $\Vc$ is the circular velocity in the solar vicinity, (i.e. $\Vc=235\kms$), and $b={r_b}/{r_{\rm{CR}}}$, with $r_b$ the bar's scale length and $r_{\rm{CR}}$ the co-rotation radius. All parameter values are taken from \citet{Chiba2022}, with $A=0.02$ and $b=0.28$. The upper row of Table~\ref{tab:bar_spipp} gives the parameters for this constant bar model. The pattern speed of the steadily rotating bar is $\Omega_{\rm{b}}=-35\kms\kpc^{-1}$ \citep{Binney2020,Chiba2021b}. Its phase angle is $\phi=28^{\circ}$ at $t=0\Gyr$, based on the azimuthal angle measured between the Sun and the major axis of the bar \citep{Wegg2015}.

\begin{table*}
	\centering
	\vspace{0.4cm}
	\caption{The parameters used for the bar and spiral arms in the model with steady rotation. $\Omega$ is in $\kms\kpc^{-1}$, and $\Vc$ in $\kms$. $r_{\rm{CR}}$, $R_{\rm{s}}$ and $h_{\rm{s}}$ are in $\kpc$. $\phi_{\rm{b}}$ and $\phi_0$ denote the initial phase angles of the bar and the spiral arms respectively. The unit of $\Sigma_0$ is $\msun\kpc^{-2}$.}
	\label{tab:bar_spipp}
	\begin{tabular}{lccccccc} % four columns, alignment for each
		\hline
		Bar  & $\Omega_{\rm{b}}$ & $A$ & $\Vc$ & $b$ & $r_{\rm{CR}}$  & $\phi_{\rm{b}}$\\
	
		Values & -35 & 0.02 & 235 & 0.28 & 6.7 & $28^{\circ}$\\
            \hline
		Spiral arm &  $\Omega_{\rm{p}}$ & $R_{\rm{s}}$ & $h_{\rm{s}}$ & $N$ & $\alpha$ & $\phi_0$  &  $\Sigma_0$\\
		
		Values &   -18.9 &1.0 & 0.1 & 2 & $9.9^{\circ}$ & $26^{\circ}$ & $2.5\times10^9$\\
		\hline
	\end{tabular}
	\vspace{-0.cm}
\end{table*}

\begin{figure*}
    \centering
     \includegraphics[width=\linewidth]{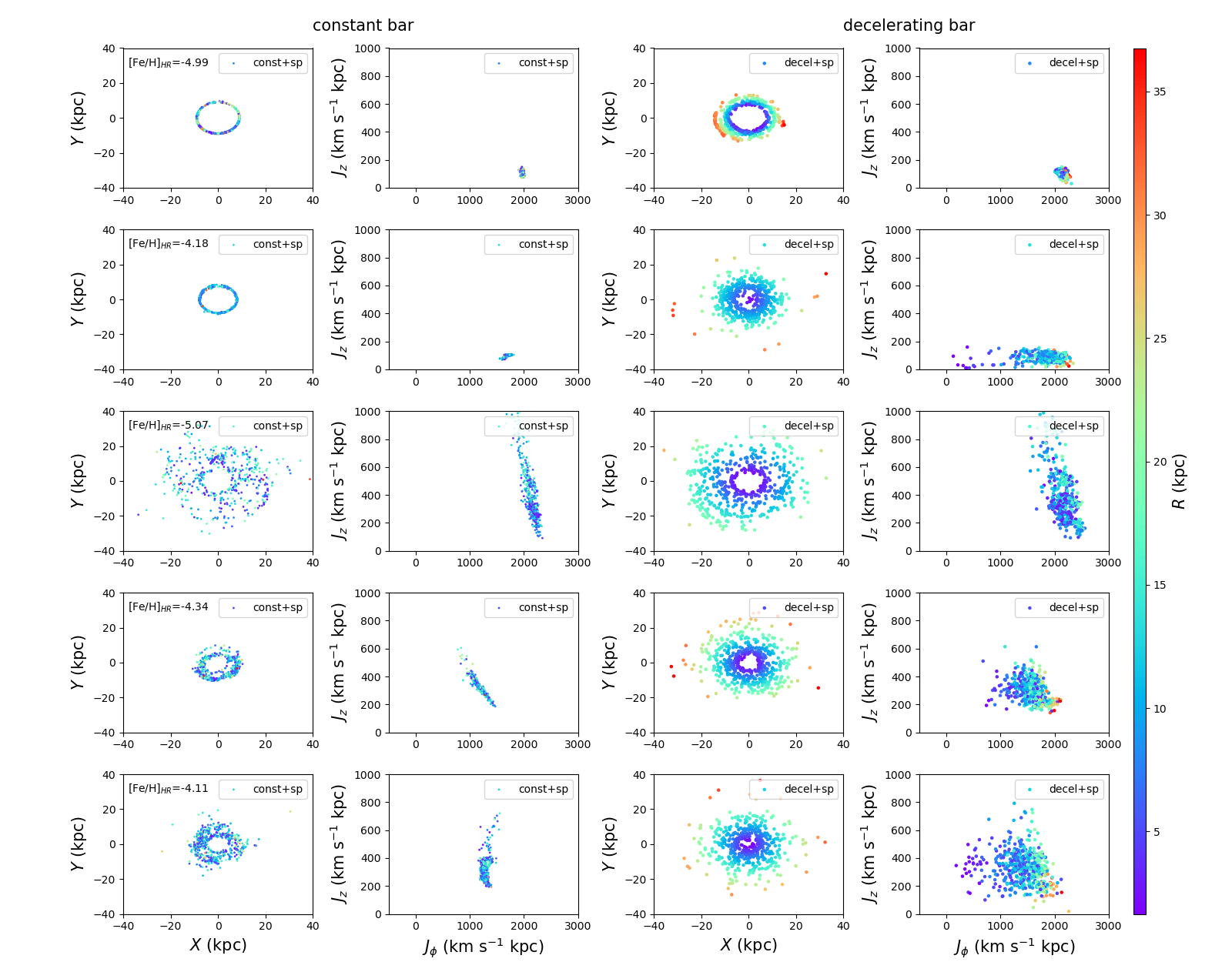}
    \caption{Distribution of the 500 sampled particles of five UMP planar stars in the ($X$, $Y$) and ($J_{\phi}$, $J_z$) space color-coded by their planar radius ($R$) 6 $\Gyr$s ago. The two left-hand columns show that most of the particles approximately maintain their orbits under a constant bar for 6 $\Gyr$s. The two right-hand columns show a much larger scatter in both of the spaces for a rapidly decelerating bar.}
    \label{fig:action_ump}
\end{figure*}

The second bar model we consider has a large initial pattern speed, which then decreases with time. The bar model is adopted from \citet{Sormani2022}, with the initial pattern speed $\Omega_{\rm{b}}=-88\kms\kpc^{-1}$ at $t=-6\Gyr$. It immediately starts to decrease steadily and reaches $\Omega_{\rm{b}}=-38\kms\kpc^{-1}$ at the present time. Meanwhile, the mass and radial profile of the bar are set to increase by factors of 2.0 and 1.2 respectively, which roughly simulates the growth of the bar. The pattern speed decreases by about $\sim55\%$ during the simulation, which is much larger than the lower limit ($\sim24\%$) constrained by the $Gaia$ DR2 RVS catalogue \citep{dr3rvs} in a similar setting \citep{Chiba2021b}. We are thus exploring the impact of an extreme decelerating bar model here. 

The spiral arms are described by a two-arm model based on \citet{Cox2002},

\begin{equation}
    \Phi_{\rm{s}}(R,\phi,z)=-4\pi G \Sigma_0\mathrm{e}^{-R/R_{\rm{s}}}\sum_n\frac{C_n}{K_n\,D_n}\!
    \cos n\gamma \big[\cosh\big(\tfrac{K_nz}{\beta_n}\big)\big]^{-\beta_n},
    \label{eq:spiralarm}
\end{equation}
where $(R,\phi,z)$ are cylindrical coordinates, and $\Sigma_0$ is the central surface density. $C_{n=1,2,3}$ are ${C_1=8/3\pi,C_2=1/2,C_3=8/15\pi}$ and represent the amplitudes of the three harmonic terms. The functional parameters are
\begin{equation}
    \begin{aligned}
    &K_n\,=\,\frac{nN}{R\sin{\alpha}},\\
    &\beta_n\,=\,K_n\,h_{\rm{s}}(1+0.4K_n h_{\rm{s}}),\\
    &\gamma\,=\,N\bigg[\phi\,-\,\frac{\ln{(R/R_{\rm{s}})}}{\tan{\alpha}}\,-\,\Omega_{\rm{p}} t\,-\,\phi_0 \bigg],\\
    &D_n\,=\, \frac{1}{1+0.3K_nh_{\rm{s}}}\,+\,K_nh_{\rm{s}},
    \end{aligned}
    \label{eq:spiral_paras}
\end{equation}
with $N$ the number of arms, $h_{\rm{s}}$ the scale height, $\alpha$ the pitch angle, $\phi_0$ the phase, and $n=1,2,3$ the three harmonic terms. 

The parameters used in the spiral arm potential can be found in the lower panel of Table~\ref{tab:bar_spipp}, most of which are adopted from \citet[]{monari2016a,monari2016b} and represent a tightly wound spiral pattern. The phase angle is $\phi_0=26^{\circ}$ at $t=0\Gyr$ \citep{monari2016b} and the pattern speed is set to be $\Omega_{\rm{p}}=-18.9\kms\kpc^{-1}$ \citep{monari2016b}. 

We perform simulations with four different perturbation setups: (i) constant bar only, (ii) constant bar plus spiral arms, (iii) decelerating bar only, and (iv) decelerating bar plus spiral arms. With these, we aim to compare the different behaviors of the sampled particles under different perturbation potentials.

% \AS{we now from N-Body simulations that spiral arms do also evolve in time, maybe add some elements to justify why it is not relevant to include this in the present study} We start the simulation by evolving the pseudo stars backwards in this potential for 6 $\Gyr$s with the trajectories stored every 0.02 $\Gyr$. 

% Then we compute the DFs for all the sampled particles at $\rm{T}\,=\,-6\,\Gyr$ with the DF model used to construct the axisymmetric potential \AS{what do you mean by compute the DF? It sounds more like a likelihood to me}. 

\begin{figure*}
    \centering
    \includegraphics[width=\linewidth]{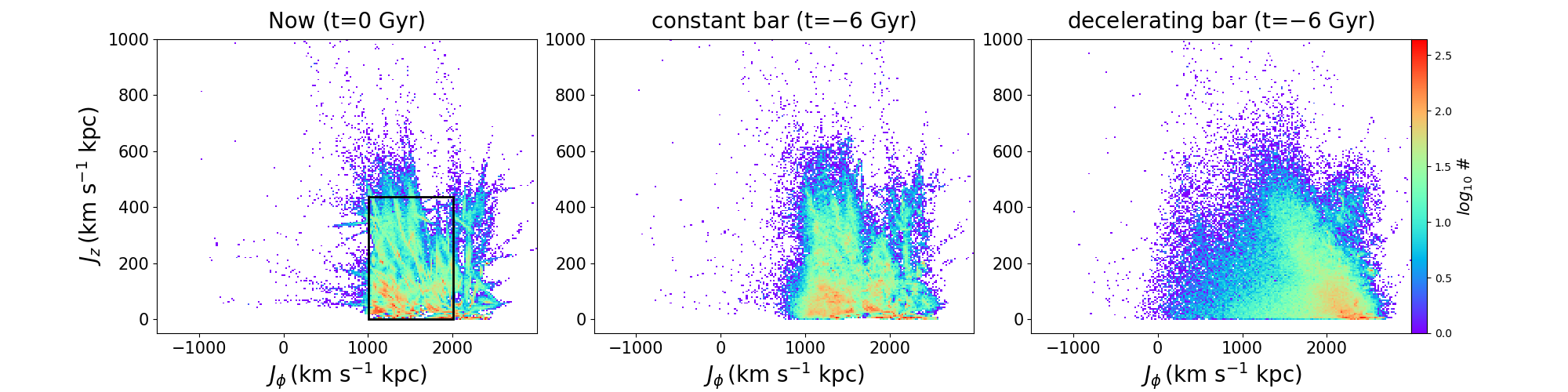}
    \caption{Density plot of all sampled particles in the action space ($J_{\phi}$, $J_{z}$). Left: the original sample of stars currently in the selection box (black rectangle): 0.5 $\leq$ $J_{\phi}/J_{\phi,\sun}$ $\leq$ 1.0, 0.5 $\leq$ $J_{z}/J_{z,\sun}$ $\leq$ 1250, with $J_{\phi,\sun}$ = 2009.92 $\kms\kpc$ and $J_{z,\sun}$ = 0.35 $\kms\kpc$. Middle: The sampled particles 6 Gyrs ago in model (ii) of a constant bar with spiral arms. They remain similar to their initial distribution in the left-hand panel. Right: The particles in model (iv) of a rapidly decelerating bar with spiral arms. They have a more extended distribution $J_{\phi}$ $6\Gyr$ ago in this model. Some low-$J_{\phi}$ particles ($J_{\phi}$ $\lesssim$ 1000 $\kms\kpc$) have gained rotation from the bar but represent only a small fraction of the entire sample.}
    \label{fig:action}
\end{figure*}

\section{Results}{\label{sec:res}}

 In Fig.~\ref{fig:action_ump}, we first show, for all five prograde planar UMP stars, the distribution after the $6\Gyr$ backwards integration of the 500 particles drawn from the uncertainties of their phase-space parameters. The different panels show both these distributions in the Galactocentric ($X,Y$) space and in the ($J_{\phi},J_{z}$) action space and all particles are color-coded by their planar radii, $R = \sqrt{X^2+Y^2}$. The two left-hand columns displays these distributions for the model with a constant bar and spiral arms. The stars mainly preserve their orbits in the spatial space and have $J_{\phi}$ spreads that remain small ($<500 \kms\kpc$). In contrast, the distributions under a decelerating bar with spiral arms (two right-hand columns) are much more scattered in both spaces. The particles typically have a large range of $J_{\phi}$ ($\sim 1000 \kms\kpc$) after the backwards integration.

We then compare the results from our experiments by showing, in Fig.~\ref{fig:action}, the density plot of all sampled particles in the action space ($J_{\phi}$, $J_{z}$). The majority of them at present (left panel) reside well within the selection box described in Sec. \ref{sec:data}. The middle panel presents the distribution of these particles 6 $\Gyr$s ago under a constant bar with spiral arms (model ii) and displays only small changes from their current properties. In the case of a rapidly decelerating bar with spiral arms (model iv), shown in the right-hand panel, we clearly see a wider distribution of $J_{\phi}$, which extends below $J_{\phi} \sim 1000 \kms\kpc$. The orbits of those sampled particles with $J_{\phi}\leq 1000 \kms$ have gained stronger rotations from the bar over the last 6 $\Gyr$s. However, their fraction is very small and the vast majority (92$\%$) of sampled particles were already rotation-dominated ($J_{\phi} \geq 1000 \kms$) 6 Gyr ago. To make a rough estimate of the fraction of stars that were in the inner Galaxy with little rotation, we use a cut of $J_{\phi}\leq 600 \kms\kpc$ to select original bulge-like orbits \citep{Binney2023c}. Only $3\%$ of the sampled particles qualify as such.

\begin{figure*}
    \centering
     \includegraphics[width=0.45\linewidth]{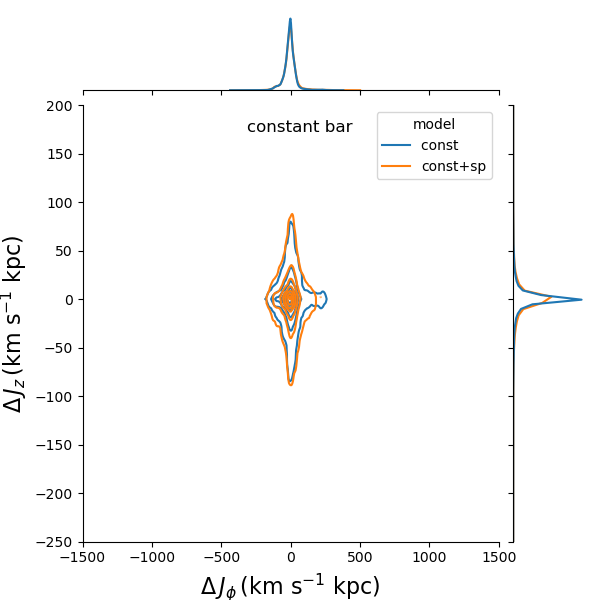}
      \includegraphics[width=0.45\linewidth]{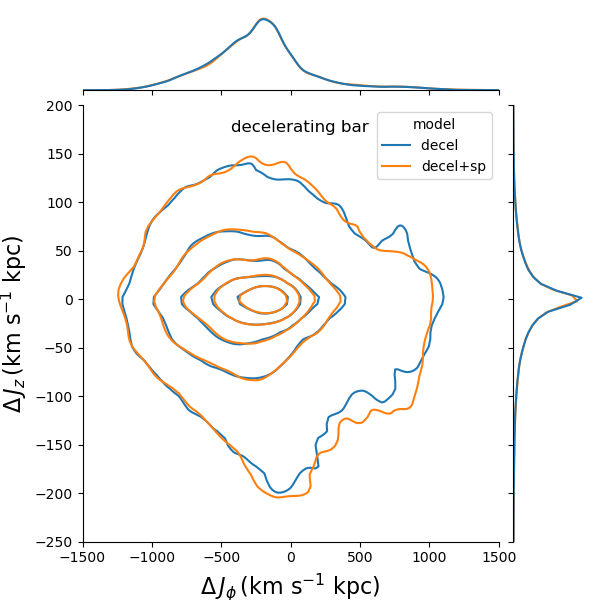}
    \caption{Density contour plot of the action changes with respect to the initial state for all sampled particles. The left-hand and right-hand panels show the case of the constant bar models and the decelerating bar models, respectively. In both panels, the orange contours represent models with spiral arms and blue contours without. Spiral arms only have a small impact on the orbital properties of the particles. The sampled stars under the decelerating bar have a much wider distribution in $\Delta J_{\phi}$ and $\Delta J_{\rm z}$ compared to those with the constant bar. The majority of the particles lose rotation ($\Delta J_{\phi}$ $<$ 0) and only a small fraction of them (19$\%$) have gained rotation from the bar during the last 6 Gyr.}
    \label{fig:daction}
\end{figure*}

We further investigate how much the different models impact the individual orbits of the sampled particles. Fig.~\ref{fig:daction} shows the density contour plot of the change in the ($\Delta\/J_{\phi}, \Delta\/J_{z}$) space for all particles under a constant bar (left-hand panel) and a rapidly decelerating bar (right-hand panel), with the changes in orbital properties defined as those at present with respect to $6\Gyr$ ago\footnote{i.e. $\Delta J = J(t=0\Gyr)-J(t=-6\Gyr)$.}. Firstly, it is clear that spiral arms have little effect on the actions of the particles: the orange contours that correspond to action changes for the models with spiral arms are very similar to those without (blue contours). In the case of a constant bar, the changes in the orbital properties are marginal, with a distribution of $\Delta J_{\phi}$ centered on $\sim 0 \kms\kpc$ and a dispersion of only $\sim 70 \kms\kpc$ (interval between the 16th and 84th percentile of the distribution of $\Delta J_{\phi}$). The dispersion under a rapidly decelerating bar is much wider: from $-$560 $\kms\kpc$ (16th percentile) to 41 $\kms\kpc$ (84th percentile). The majority of the particles in fact lose rotation  within the $6\Gyr$ of the simulation and only a small fraction of them (19$\%$) gain rotation from interactions with the bar. This effect could be related to the migration of the bar's corotation resonance-trapped regions \citep[see details in][]{li2023b}. A bar that changes in pattern speed also changes its corotation with time, yielding much more efficient radial migration than in the case of a constant bar \citep[see \eg][]{monari2016a}. Secondly, 46$\%$ of the sampled particles have positive $\Delta J_{z}$ indicating that nearly half of the stars have become kinematically hotter and gained vertical motion as their orbits are perturbed by the bar. 

\section{Discussions}{\label{sec:dis}}

In this work, we explore the possible origin of the low-metallicity prograde planar stars found in the solar vicinity (heliocentric distances $\lesssim 5\kpc$) by running orbital integrations backwards for $6\Gyr$ under different bar models. The results show that a rotating bar cannot be a robust mechanism to explain the existence of these observed stars. First, a constantly rotating bar has little impact on the orbits of the stars. In an extreme case of a rapidly decelerating bar, some of these stars can be trapped in the corotation resonance region and be shepherded from the inner Galaxy to the solar neighborhood. However, the majority of the sampled particles (92$\%$) were already rotation-dominated ($J_{\phi}$ $\geq$ 1000 $\kms\kpc$) 6 Gyr ago. The chance of them starting with small rotation is very low ($< 3\%$ with $J_{\phi} \lesssim 600 \kms\kpc$). These old prograde planar stars that are currently present in the solar neighborhood possibly have varied origins, as tentatively shown by the initial analysis of their chemical abundances \citep{dovgal23}. Most of them start with rotation-dominated orbits after their birth and thus were either born in-situ in the proto-MW disc, came from accreted systems that merged onto the MW with very prograde orbits, or were brought in with the clumps that formed the proto-MW \citep{sestito21}.

From the modeling aspect, our method is capable of exploring the origins of stars by tracing them under different bar models. However, there are key limitations to this approach. Firstly, the decelerating bar model is only a toy model that cannot represent the true evolution history of the bar in the Galaxy. For example, the pattern speed drops drastically faster than for current values estimated in the Galaxy \citep{Chiba2021a}. Secondly, the test-particle simulation method does not include any response of the stellar systems to the perturbations by the bar and the spiral arms that is due to the self-gravity of the system itself. Thirdly, the method does not take into account the evolution/increase of the background potential of the Galaxy itself over the last 10 Gyr, especially in the epochs between 6 and 10 Gyr ago, and it does not take into account the possibility that recurring spiral arms with resonances at different radii, overlapping with the bar's resonances also at different radii over time \citep{sellwood02, minchev10}, could enhance the migration process of the old stars once the process has started. Future improvements of this method need more explicit knowledge of the evolution history of the bar's pattern speed and radial profiles. The toy models presented here are nevertheless useful to explore possible scenarios before moving on to more complex modeling and simulations.

From the observational side, the strong selection effect of different ground-based survey samples used in this work may lead to misunderstanding their true distribution. Any quantitative interpretation will require to determine a comprehensive selection function for the data used. This will be greatly facilitated by systematic surveys of low-metallicity stars, such as the upcoming WEAVE \citep{jin2023} and 4MOST surveys \citep{4most}. In addition, the ability to detect the very low-metallicity prograde planar stars is still mainly limited to lines of sights away from the disk, towards the Galactic caps, as the search for these stars in the disc regions is made difficult by the overwhelming population of more metal-rich stars and by increasingly high extinction. Future near-infrared astrometric surveys, such as the MOONS survey \citep{moons} and $GaiaNIR$ \citep{gaiair} would certainly be a significant improvement but will require new techniques to identify the most metal-poor stars that are currently being discovered in photometric surveys using optically blue wavelengths.

% We still need to find a way to find these stars  the Near-Infra-Red (NIR) astrometry \citep{gaiair}, which is able to penetrate obscured regions and measure the motions of stars towards the Galactic center.

\begin{acknowledgements}

ZY, NFM, BF, GM, and RAI acknowledge funding from the European Research Council (ERC) under the European Unions Horizon 2020 research and innovation programme (grant agreement No. 834148). CL, AS, BF, GM, VH and GK acknowledge funding from the ANR grant N211483 MWdisc. ZY and NFM gratefully acknowledge support from the French National Research Agency (ANR) funded project ``Pristine'' (ANR-18-CE31-0017). AAA acknowledges support from the Herchel Smith Fellowship at the University of Cambridge and a Fitzwilliam College research fellowship supported by the Isaac Newton Trust. ES acknowledges funding through VIDI grant "Pushing Galactic Archaeology to its limits" (with project number VI.Vidi.193.093) which is funded by the Dutch Research Council (NWO). This research has been partially funded from a Spinoza award by NWO (SPI 78-411). This research was supported by the International Space Science Institute (ISSI) in Bern, through ISSI International Team project 540 (The Early Milky Way). ZY thanks the discussions with Adam Dillamore and Vasily Belokurov during the MW-Gaia workshop supported by COST Action CA18104: MW-Gaia.

This work has made use of data products from the Guo Shou Jing Telescope (LAMOST). LAMOST is a National Major Scientific Project built by the Chinese Academy of Sciences. Funding for the project has been provided by the National Development and Reform Commission. LAMOST is operated and managed by the National Astronomical Observatories, Chinese Academy of Sciences.

This work has made use of data from the European Space Agency (ESA) mission \Gaia\ (\url{https://www.cosmos.esa.int/gaia}), processed by the \Gaia\ Data Processing and Analysis Consortium (DPAC, \url{https://www.cosmos.esa.int/web/gaia/dpac/consortium}). Funding for the DPAC has been provided by national institutions, in particular the institutions participating in the \Gaia\ Multilateral Agreement.

\end{acknowledgements}



\bibliographystyle{aa}
\bibliography{example}

%%%%%%%%%%%%%%%%%%%%%%%%%%%%%%%%%%%%%%%%%%%%%%%%%%

%%%%%%%%%%%%%%%%% APPENDICES %%%%%%%%%%%%%%%%%%%%%

\appendix

%%%%%%%%%%%%%%%%%%%%%%%%%%%%%%%%%%%%%%%%%%%%%%%%%%

\end{document}